%% file: main.tex
\PassOptionsToPackage{dvipsnames}{xcolor}
\documentclass[a4paper,UKenglish,english,cleveref,autoref,thm-restate]{lipics-v2021}
\input{preamble}

\title{Dandelion: Certified Approximations of Elementary Functions}

\author{Heiko Becker}{MPI-SWS, Saarland Informatics Campus (SIC), Germany}{hbecker@mpi-sws.org}{}{}
\author{Mohit Tekriwal}{University of Michigan -- Ann Arbor, USA}{tmohit@umich.edu}{}{}
\author{Eva Darulova}{Uppsala University, Sweden}{eva.darulova@it.uu.se}{}{}
\author{Anastasia Volkova}{Nantes Universit\'{e}, France}{anastasia.volkova@univ-nantes.fr}{}{}
\author{Jean-Baptiste Jeannin}{University of Michigan -- Ann Arbor, USA}{jeannin@umich.edu}{}{}
\authorrunning{H. Becker, M. Tekriwal, E. Darulova, A. Volkova, and J-B. Jeannin} %mandatory. First: Use abbreviated first/middle names. Second (only in severe cases): Use first author plus 'et al.'

\Copyright{Heiko Becker, Mohit Tekriwal, Eva Darulova, Anastasia Volkova and Jean-Baptiste Jeannin}

 %TODO mandatory; please add comma-separated list of keywords

\funding{This work was supported in part by the University of Michigan.}%optional, to capture a funding statement, which applies to all authors. Please enter author specific funding statements as fifth argument of the \author macro.

\acknowledgements{\input{sections/acknowledgement}}%optional

\nolinenumbers %uncomment to disable line numbering

\EventEditors{June Andronick and Leonardo de Moura}
\EventNoEds{2}
\EventLongTitle{13th International Conference on Interactive Theorem Proving (ITP 2022)}
\EventShortTitle{ITP 2022}
\EventAcronym{ITP}
\EventYear{2022}
\EventDate{August 7--10, 2022}
\EventLocation{Haifa, Israel}
\EventLogo{}
\SeriesVolume{237}
\ArticleNo{29}
\begin{document}

\maketitle

%TODO mandatory: add short abstract of the document
\begin{abstract}
  \input{sections/abstract}
\end{abstract}

\input{sections/intro}
\input{sections/overview}
\input{sections/taylorApprox}
\input{sections/sturmSeq}
\input{sections/implementation}
\input{sections/evaluation}
\input{sections/related}
\input{sections/conclusion}
\bibliography{references.bib}

\end{document}

%% file: preamble.tex
\usepackage{hyperref}
\usepackage{todonotes}
\usepackage{xspace}
\usepackage{caption}
\usepackage{subcaption}
\usepackage{booktabs}

\usepackage[scaled=0.8]{beramono}  % our monospace font
\usepackage[T1]{fontenc}  % this is necessary for beramo to work

\definecolor{keywordsColor}{rgb}{0.000000, 0.000000, 0.635294}

\lstdefinelanguage{scala}{
  alsoletter={@=>},
  morekeywords={
    fun, let, val, in, end, if, then, else, case, of, def, opt, noopt, and},
  sensitive=true,
  morecomment=[l]{//},
  morecomment=[s]{(*}{*)},
  commentstyle=\color{gray},
  showstringspaces=false,
  columns=fullflexible,
  mathescape=true,
  numberstyle=\tiny,
  basicstyle=\ttfamily\small,
  backgroundcolor=\color{white},
  numbersep=5pt,
  stepnumber=2,
  numbers=none,                   % where to put the line-numbers
  morestring=[b]``,
  keywordstyle=\color{keywordsColor}\bfseries
}

\lstdefinelanguage{Hol}{
  morekeywords=[1]{val, Define, prove},
  morekeywords=[2]{Type, Prop, Set, true, false, option},
  morecomment=[s]{(*}{*)},
  sensitive=true,
  morecomment=[l]{//},
  morecomment=[s]{(*}{*)},
  commentstyle=\color{gray},
  showstringspaces=false,
  columns=fullflexible,
  mathescape=true,
  numberstyle=\tiny,
  basicstyle=\ttfamily\small,
  backgroundcolor=\color{white},
  numbersep=5pt,
  stepnumber=2,
  numbers=none,                   % where to put the line-numbers
  morestring=[b]``,
  keywordstyle=\color{keywordsColor}\bfseries
}
\newcommand{\Sin}{\ensuremath{\sin}}
\newcommand{\Cos}{\ensuremath{\cos}}
\newcommand{\Exp}{\ensuremath{\exp}}
\newcommand{\Atn}{\ensuremath{\tan^{-1}}}

\newcommand{\Log}{\ensuremath{\ln}}

\newcommand{\eg}{\textit{e.g.}\xspace}
\newcommand{\ie}{\textit{i.e.}\xspace}

\newcommand{\etal}{\textit{et al.}\xspace}

\newtheoremstyle{thmnonit}% <name>
{}% <Space above, use default>
{}% <Space below, use default>
{\normalfont}% <Body font>
{}% <Indent amount>
{\bfseries}% <Theorem head font>
{.}% <Punctuation after theorem head>
{.5em}% <Space after theorem head>
{}% <Theorem head spec (can be left empty, meaning `normal')>
\theoremstyle{thmnonit}

\newtheorem{thm}{Theorem}

\addto\extrasenglish{%

}

%% file: sections/acknowledgement.tex
The authors would like to thank John Harrison for the insightful
discussion and for providing the source code for his paper that inspired the
Dandelion work.
Further, we thank Magnus Myreen and Michael Norrish for their help with improving
the HOL4 implementation of Dandelion.
We also thank Samuel Coward for helping us with the MetiTarski evaluation.
Finally, we thank the anonymous ITP reviewers for their feedback on the
paper.
%%% Local Variables:
%%% mode: latex
%%% TeX-master: "../main"
%%% End:

%% file: sections/abstract.tex
Elementary function operations such as \Sin{} and \Exp{} cannot in general be % generally
computed exactly on today's digital computers, and thus have to be
approximated. The standard approximations in library functions typically provide
only a limited set of precisions, and are too inefficient for many applications.
Polynomial approximations that are customized to a limited input domain and
output accuracy can provide superior performance. In fact, the Remez algorithm
computes the best possible approximation for a given polynomial degree, but has
so far not been formally verified.

This paper presents \emph{Dandelion}, an automated certificate checker for
polynomial approximations of elementary functions computed with Remez-like
algorithms that is fully verified in the HOL4 theorem prover.
Dandelion checks whether the difference between a polynomial approximation and
its target reference elementary function remains below a given error bound for
all inputs in a given constraint.
By extracting a verified binary with the CakeML compiler, Dandelion can validate
certificates within a reasonable time, fully automating previous manually
verified approximations.
%%% Local Variables:
%%% mode: latex
%%% TeX-master: "../main"
%%% End:

%% file: sections/intro.tex
% !TEX root = ../main.tex
\section{Introduction}

Exact computation in real-number arithmetic is in general too inefficient for
most applications~\cite{Boehm2019}, and is thus typically replaced by
finite-precision (floating-point or fixed-point) arithmetic.
While arithmetic operations such as addition and multiplication are
well-supported and efficient, real-world code often also needs to support
elementary functions such as \Sin{} and \Exp{}. Such functions cannot be
computed exactly on today's digital hardware and thus necessarily have to be
approximated.
For floating-point arithmetic, libraries provide general-purpose approximations
for a limited set of formats, e.g. correctly rounded to single or double
precision~\cite{lim2022one,daramy2003cr}. However, these can be inefficient for applications that do
not need quite as much accuracy, or that only need to work for a limited set of
inputs~\cite{DarulovaV19,kupriianova2014metalibm}. Furthermore, many applications, especially in the embedded systems
domain, operate with fixed-point arithmetic, for which efficient library
approximations do not exist~\cite{izycheva2019synthesizing}.

When library function implementations are suboptimal, a possible solution is to
generate approximations of elementary functions on demand with custom
accuracy---exactly the accuracy that is needed by the application and its
context.
Indeed, automated algorithms exist for generating polynomial approximations of
elementary functions with a given polynomial degree or given bound on the
\emph{approximation error}. For instance, Remez-like
algorithms~\cite{pachon2009barycentric} generate the best polynomial
approximation, i.e. the one with the smallest approximation error for a given
polynomial degree.
We can, for example, approximate $\Exp$ on $[0, 0.5]$ by $ 0.999 + 1.001 \times x +
0.484 \times x^{2} + 0.215 \times x^{3}$ with an approximation error of $2.63\times10^{-5}$.

Remez-like algorithms are used in several automated tools for generating custom
approximations~\cite{Sollya,kupriianova2014metalibm}, however, these
implementations are not formally verified. This is especially problematic
because the algorithms are generally tricky to get
right~\cite{muller2006elementary}\footnote{Muller warns: ``[..] even if the
outlines of the [Remez] algorithm are reasonably simple, making sure that an
implementation will always return a valid result is sometimes quite difficult.''
(\cite{muller2006elementary}, page 52).}.

%This paper, mini-contributions
In this paper, we implement and prove correct \emph{Dandelion}, a fully
automated and formally verified certificate checker for polynomial
approximations computed by Remez-like algorithms.
Dandelion is implemented and fully verified inside the HOL4 theorem
prover~\cite{HOL4Tutorial}.
A certificate for Dandelion consists of an elementary function $f$, an input
interval $I$, and a polynomial approximation $p$ and an approximation error $\epsilon$ returned by a
Remez-like algorithm.
We prove once and for all the correctness theorem that if Dandelion returns true
for a certificate, the encoded error is a true upper bound to the
difference between the elementary function and the encoded polynomial:
$\max_{x \in I} | f (x) - p(x) | \le \epsilon$.

Dandelion's certificates are minimal, requiring only inputs and outputs of the
approximation algorithm to be recorded.
Additionally, Dandelion certifies the known best possible approximations of
Remez-like algorithms, together, making it widely applicable.
Previous work focused on manual proofs~\cite{harrison1997verifying};
certifies only results of Chebyshev approximations, which are not as accurate
as those computed by Remez-like algorithms~\cite{brehard2019certificate};
or their verification-technique is mainly based on interval arithmetic~\cite{martin2016coqinterval}.
Dandelion is the first tool that automates the approach of Harrison~\cite{harrison1997verifying},
and thus the key challenge that Dandelion solves is \emph{automation};
Dandelion requires no user interaction, making it the first fully automated
validator for results of Remez-like algorithms based on polynomial zero finding.

One may think that verifying an implementation of a Remez-like algorithm
should be favored over validating each run separately.
However, correctness proofs for one implementation generally do not apply to other implementations,
and thus would have to be re-done with every change.
In contrast, by certifying only the end-result, Dandelion is indifferent
to the implementation choices and thus immediately more widely applicable.

Harrison~\cite{harrison1997verifying} has manually verified a polynomial
approximation of the exponential function in HOL-Light~\cite{hollight}.
The methodology presented is general, but was never automated.
Dandelion borrows the high-level approach from Harrison's manual proof,
automating the key ideas to validate results of Remez-like algorithms.

While the idea of automating an existing development may seem simple, we faced
two major challenges to make automated certification practical.
First, computations in theorem provers are generally slower than those in unverified
tools, making certain designs impractical.
Second, some definitions of Harrison use non-computable functions and thus cannot be used
in an automated approach.
To speed-up the computations, we extract Dandelion as a verified binary using
the CakeML compiler~\cite{CakeML}.
The extracted binary enjoys the same correctness guarantees as our in-logic
implementation, and makes checking a certificate fast: a single certificate is
checked on average within 6 minutes.
We overcome the problem of non-computable functions by identifying computable
versions and proving equivalence between the computable and non-computable
functions.

Dandelion can be used as a verifier for any Remez-like algorithm.
In our evaluation, we use Dandelion to certify a number of approximations
generated from FPBench~\cite{fpbench} and the work by Izycheva
\etal{}~\cite{izycheva2019synthesizing}.
Our evaluation shows that certificate checking in Dandelion is fast, and that
Dandelion certifies, for an elementary function $f$ and polynomial $p$,
approximation errors on the same order of magnitude as the infinity norm ($\max_{x \in I(x)} | f (x) - p(x) |$).
We also encode the original proof-goal of Harrison as a certificate---Dandelion
reduces its proof to a single line of code.

\paragraph*{Contributions}
In summary, this paper provides the following contributions:
\begin{itemize}
\item a HOL4 implementation of Dandelion\footnote{The source code of Dandelion is publicly available at \url{https://github.com/HeikoBecker/Dandelion}.}, a verified certificate checker for polynomial approximations
\item a verified binary extracted using CakeML to make certificate checking fast, and
\item an evaluation of Dandelion's performance on a set of benchmarks, comparing it with the state-of-the-art.
\end{itemize}
%%% Local Variables:
%%% mode: latex
%%% TeX-master: "../main"
%%% End:

%% file: sections/overview.tex
% !TEX root = ../main.tex
\section{Overview}\label{sec:overview}

\begin{figure}[t]
  \begin{subfigure}[t]{.49\textwidth}
    \begin{lstlisting}[escapechar=\%,basicstyle=\footnotesize\ttfamily,numbers=left,numberstyle=\tiny,language=scala]
def polToCart_x(radius: Real,
  theta: Real): Real = {
 require(((1.0 <= radius) &&
  (radius <= 10.0) && (0.0 <= theta) &&
  (theta <= 360.0)))
 val pi = 3.14159265359
 val radiant = (theta * (pi / 180.0))
 (radius * cos(radiant))
}
    \end{lstlisting}
    \vspace{1.4em} %For alignment
    \caption{Example kernel using a elementary function}\label{subfig:ex1_scala}
  \end{subfigure}
  \begin{subfigure}[t]{.49\textwidth}
    \begin{lstlisting}[escapechar=\%,basicstyle=\footnotesize\ttfamily,numbers=left,numberstyle=\tiny,language=scala]
def polToCart_x(radius: Fixed,
 theta: Fixed): Fixed = {
 val pi = 3.14159265359
 val radiant = (theta * (pi / 180.0))
 val _tmp = (1.3056366443634033 +
  (radiant * (-1.2732393741607666 +
  (radiant * (0.2026423215866089 +
  (radiant * 3.3222216089257017e-09))))))
 (radius * _tmp)
  }
    \end{lstlisting}
    \caption{Example with polynomial approximation for $\Cos$}\label{subfig:ex1_res}
  \end{subfigure}
  \begin{subfigure}[c]{.99\textwidth}
\begin{lstlisting}[escapechar=\%,basicstyle=\footnotesize\ttfamily,numberstyle=\tiny]
cos_cert = <| f := Fun Cos (Var "radiant"); n := 32;
  (* p (x) ~ 1.305 - 1.273 * x + 0.202 * %{\color{gray}$x^2$}% + 3.322 * %{\color{gray}$10^{-9}$}% * %{\color{gray}$x^3$}% *)
  p := [5476237/4194304; -5340353/4194304; 1699887/8388608; 3740489/1125899906842624];
  %$\varepsilon$% := 7661335245848499811609873770389478739611431267987/(25 * 10^48); (* ~0.306 *)
  I := [("radiant", (0, 314159265359/50000000000))]; (* ~ x in [0, 6.284] *) |>
\end{lstlisting}
\caption{Certificate for the approximation of $\Cos$ in the example}\label{subfig:ex1_cert}
  \end{subfigure}
\caption{Example kernel using a elementary function (top-left), the kernel with
a polynomial approximation (top-right), and the certificate for Dandelion
(bottom)}\label{fig:ex1}
\end{figure}

Before we dive into the technical details of Dandelion, we give an overview
of our toolchain and the proofs that Dandelion performs automatically using the
example in~\autoref{fig:ex1}.
The starting point is the code in~\autoref{subfig:ex1_scala} that converts polar to
cartesian coordinates, and returns the resulting $x$ component.
This code could for example be part of an autonomous car or a drone, and
inaccuracies in the conversion of coordinates may have catastrophic
effects~\cite{munoz2015daidalus}.
Chip sizes and energy budgets in these devices are usually small, and thus
using a fully-fledged floating-point unit is not always possible.
As an alternative, code is often implemented in fixed-point arithmetic, which,
however, does not come with standard and efficient library implementations
of elementary functions~\cite{izycheva2019synthesizing}.
Hence, an engineer may approximate the function $\Cos$ on line 8
in~\autoref{subfig:ex1_scala} with a custom polynomial approximation shown in
\autoref{subfig:ex1_res}, for instance using the state-of-the-art synthesis tool
Daisy~\cite{izycheva2019synthesizing}\footnote{Daisy can also synthesize
suitable finite-precision types for the polynomial (not shown here), and generate certificates that formally verify the \emph{roundoff} error bound of this polynomial implementation~\cite{becker2018verified}.}.

Daisy internally calls a Remez-like algorithm to generate the polynomial
approximation of $\Cos$, but the approximation algorithm and Daisy itself are
not (formally) verified. With Dandelion, we can straight-forwardly instrument
Daisy to generate the certificate shown in~\autoref{subfig:ex1_cert} that
encodes the elementary function to be approximated (\lstinline{f}), the
approximating polynomial (\lstinline{p}), the approximation error
($\varepsilon$), and the range on which the approximation is supposed to be
valid (\lstinline{I}), and an additional parameter \lstinline{n} which we
explain later.
Note that the input interval \lstinline{I} recorded in the certificate captures
the direct inputs to the elementary function $\Cos$ and is thus different from
the input interval in the \lstinline{require} clause that captures inputs to the
overall function \lstinline{polToCart_x}.
Dandelion validates this certificate in 31 seconds and  proves the HOL4 theorem
\begin{thm}{}\label{thm:ex1_thm}:
  $\forall x.\;x \in$ \lstinline{I}$(x) \Rightarrow | \Cos (x) - $\lstinline{p}$(x) | \leq \varepsilon$
\end{thm}
If the approximation error had not been correct, the binary would emit an error
message, explaining which part of the validation failed.

The certificate in \autoref{subfig:ex1_cert} uses only a single elementary
function.
In general, Dandelion supports more complicated elementary function expressions,
like $\Exp(x \times \frac{1}{2})$, and $\Sin(x - 1) + \Cos (x + 1)$.
Like Remez-like algorithms, we only require the functions to be
univariate, \ie{} the certificate can only have a single free variable.
Any approximation tool that can generate these certificates can be used to
generate inputs for Dandelion, and Dandelion can be used independently of a particular
approximation algorithm implementation.

The approach used by Dandelion has been laid out previously in
a manual proof for the exponential function by Harrison~\cite{harrison1997verifying} (\autoref{sec:manual-proof} overviews the main theorems and ideas).
The presented high-level approach is general, but
a key challenge that Dandelion solves is to automate each step and extend them
to more complex expressions (\autoref{sec:overview-automation}).

\subsection{Manual Proof by Harrison}\label{sec:manual-proof}
Harrison has manually verified an approximation by Tang~\cite{tang1989table}
of the exponential function, showing that:
$\forall x. x \in [- 0.010831, 0.010831] \Rightarrow | ((e^{x} - 1) - p (x) | \leq 2^{-33.2}$.
The manual verification by Harrison is split into two steps.
First, Harrison simplifies the overall proof goal to a proof about polynomials,
by replacing $e^{x}-1$ with a high-accuracy truncated Taylor series $q$.
By truncating the series after the 7th term, the approximation error of the
series becomes $2^{-58}$ and the overall proof-goal is reduced from
$| ((e^{x} - 1) - p (x) | \leq 2^{-33.2}$ with the triangle inequality to
\begin{equation}\label{eqn:Harrison1}
| q(x) - p (x) | \leq 2^{-33.2} - 2^{-58}
\end{equation}
The difference between $q(x)$ and $p(x)$ itself is a polynomial $h(x)$, and thus
this  first step reduces the overall proof goal to proving an upper bound on the
polynomial $h$.
As $h$ represents the difference between two polynomial approximations, the
points where $h$ attains its maximum value (\ie{} its extremal points) are those
where the approximation error is the largest.
It thus suffices to reason about the extremal points of $h$ for proving
the inequality.

To prove the polynomial inequality, Harrison proved two well-known mathematical
theorems in HOL-Light.
The first theorem proves that polynomial $p$ on the closed interval $[a,b]$
attains its extremal values either at the outer points or at the points where
the first derivative is zero:
\begin{thm}\label{thm:fstderiv}
Let $p$ a differentiable, univariate polynomial, defined on $[a,b]$ and $M$ a real number, then
\begin{align*}
   &|\,p (a)\,| \leq M\wedge |\,p (b)\,| \leq M\;\wedge \\
   &(\forall x.\;a\leq x \leq b\;\wedge\;p'(x) = 0 \Rightarrow |\,p (x)\,| \leq M) \Rightarrow\\
   &(\forall x.\;a \leq x \leq b \Rightarrow |\,p (x)\,| \leq M)
\end{align*}
\end{thm}
The second theorem is called \emph{Sturm's theorem} and proves that the exact
number of zeros of a polynomial can be computed from the so-called Sturm
sequence of polynomials:
\begin{thm}\label{thm:sturm}
 Let $p$ a differentiable, univariate polynomial, defined on $[a,b]$.
If $p$ has non-zero values on both $a$ and $b$, and its derivative is not the
constant zero function,
then we call \textit{Sturm} $ss$ the sturm sequence for $p$, and the
set of zeros of $p$ has size $V(a, ss) - V(b, ss)$.
\end{thm}
The Sturm sequence $ss$ of polynomial $p$ is defined recursively as
\[
  ss_{0} = p \qquad{} ss_{1} = p' \qquad{} ss_{i+1} = - \text{rem}\,(ss_{i-1}, ss_{i})
\]
where $\text{rem}$ computes the remainder of the polynomial division
$\frac{ss_{i-1}}{ss_{i}}$. Computation stops once the remainder becomes the constant $0$ polynomial.
Function $V(a, ss)$ in \autoref{thm:sturm} computes the number of sign changes
when evaluating the polynomials in the list $ss$ on value $a$.

To prove the final inequality, Harrison computes unverified guesses for both the
Sturm sequence of $h'(x)$ and the zeros of $h'(x)$ using Maple, and manually
validates them in HOL4 using \autoref{thm:sturm}.
By knowing the number of zeros, and their values, Harrison then provably
derives an upper bound on the extremal values of polynomial $h$ using
\autoref{thm:fstderiv}.

\subsection{Automated Proofs in Dandelion}\label{sec:overview-automation}
\begin{figure}[t]
  \centering
  \includegraphics[scale=.68]{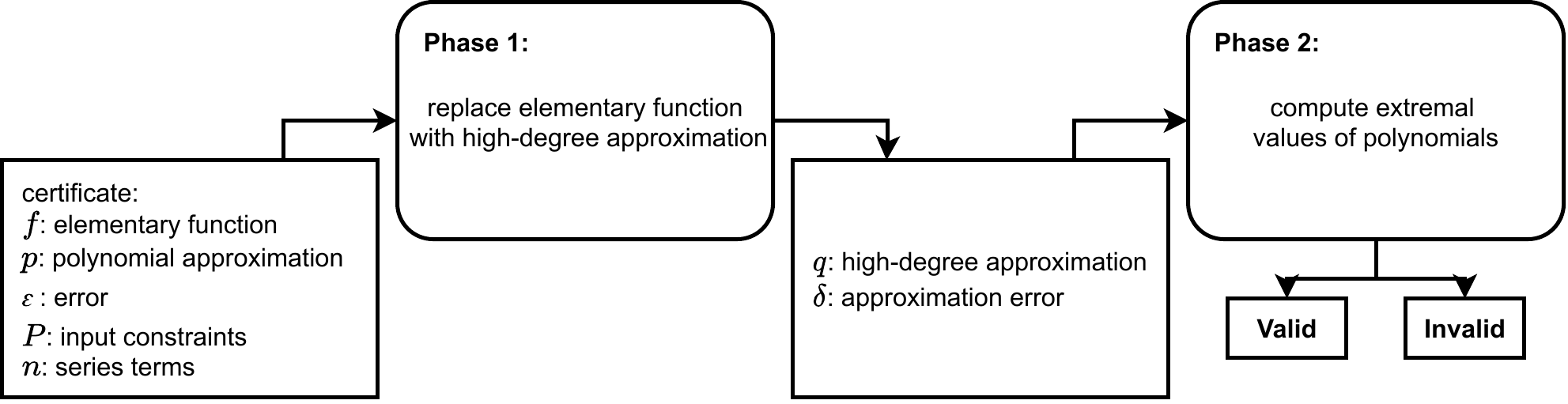}
  \caption{Overview of Dandelion toolchain}\label{fig:overview}
\end{figure}

As in Harrison's approach, Dandelion splits the proof into two parts.
In the first phase, Dandelion replaces elementary functions in the certificate
by high-accuracy approximations computed inside HOL4.
The second phase then proves that the approximation error is a correct upper bound
on the extremal points of the resulting polynomial by finding zeros of the
derivative and bounding the number of zeros with Sturm sequences.
The key differences between Dandelion and the proof by Harrison is that Dandelion
supports more elementary functions, \ie{} $\Exp{}$, $\Sin{}$, $\Cos{}$, $\Log{}$,
and $\Atn{}$, and that its certificate checking is fully automated
and does not require any user interaction or additional proofs.
\autoref{fig:overview} gives an overview of the automatic computations done
by Dandelion.
We explain them at a high-level for our example from \autoref{fig:ex1}.

To prove the overall correctness theorem (\autoref{thm:ex1_thm}), Dandelion
first computes a computable high-accuracy polynomial approximation for $\Cos$,
denoted by $q$, using a truncated Taylor series of degree \lstinline{n}, with an
approximation error of \texttt{3.77e-3}.
Generally, the certificate parameter \lstinline{n} defines the number of
series terms computed for the truncated Taylor series in Dandelion.
Exactly as for Harrison's manual proof, Dandelion proves an
upper bound on the difference between $q$ and the target polynomial
approximation $p$ (\autoref{eqn:Harrison1}).
Coming up with a general approach for computing accurate truncated Taylor series of
arbitrary elementary functions was a major challenge for Dandelion---we
implemented a library of general purpose Taylor series for the supported elementary functions.
Given a target degree, Dandelion automatically computes a polynomial implementation
and proves an approximation error for the truncated series.
We explain the first phase in more detail in~\autoref{sec:taylorApprox}.

In the second phase, Dandelion computes an upper bound on the polynomial $h(x) = q(x) - p(x)$ exactly
like Harrison, by reasoning about the zeros of its derivative $h'$, using Sturm
sequences to bound the number of zeros of $h'$.
Based on the number of zeros, Dandelion uses an (unverified) oracle to automatically
come up with a list of
zeros of $h'$.

To prove the final bound on $h$, Dandelion checks for each
zero of $h'$ that the value of $h$ at this point is smaller than or equal to the
residual error $\varepsilon - \texttt{3.77e-3}$.
Harrison's definition of Sturm sequences is defined as a non-computable
predicate, involving an existentially quantified definition of polynomial division.
Key to Dandelion is the implementation of a computable version of polynomial
division, as well as Sturm sequences, in combination with equivalence proofs relating
them to Harrison's predicates.
We explain how Dandelion computes the Sturm sequences automatically and how
Dandelion estimates zeros of polynomials in more detail in~\autoref{sec:sturmSeq}.

Computing the Sturm sequence is the most computationally expensive part of Dandelion,
which we found to be impractical to do in logic.
Thus we extract a verified binary using the CakeML compiler for the second phase
of Dandelion only.
For the extraction to work, we translated the HOL4 definitions of the second
phase into CakeML source code via CakeML's proof-producing synthesis tool~\cite{JAR20}.
We explain the extraction with CakeML in~\autoref{sec:implementation}.
%%% Local Variables:
%%% mode: latex
%%% TeX-master: "../main"
%%% End:

%% file: sections/taylorApprox.tex
% !TEX root = ../main.tex
\section{Automatic Computation of Truncated Taylor Series}\label{sec:taylorApprox}
As in Harrison's manual proof, Dandelion replaces in a first step the
elementary function in the certificate by a high-accuracy polynomial
approximation.
The crucial difference is that Dandelion automates all of the manual steps, which we
explain next.

When checking a certificate for function $f$, with input range $I$, Dandelion automatically replaces
every occurrence of an elementary function in $f$ with a truncated Taylor series
$t_{f,n}$.
Below, we take $f$ to be a single elementary function and will discuss the extension
to more complicated elementary function expressions later.
The parameter $n$ of $t_{f,n}$ is part of the certificate and specifies the
number of terms computed for the truncated series, \ie{} if $n$ is 32,
Dandelion truncates the Taylor series of $f$ after the 32nd term.
The final result of the phase is a high-accuracy polynomial approximation of
function $f$, $q_{f,n}$, and an overall approximation error $\delta_{f,n}$.
For the simple case where $f$ is a single elementary function, $q_{f,n}$ and
$t_{f,n}$ are the same.
Once we extend Dandelion to more complicated expressions, Dandelion
combines different instances of $t_{f,n}$, into the final $q_{f,n}$, as we
explain later.
We implement the first phase in a HOL4 function \lstinline{approxAsPoly} and
prove soundness of \lstinline{approxAsPoly} once and for all in HOL4:
\begin{thm}[First Phase Soundness]\label{thm:approxPolySound}
  \begin{align*}
    {\normalfont \texttt{approxAsPoly}}\;f\;I\;n = \texttt{Some} (q_{f,n}, \delta_{f,n}) \Rightarrow
    (\forall x.\;x \in I \Rightarrow |\,f (x) - q_{f,n} (x)\,| \leq \delta_{f,n})
  \end{align*}
\end{thm}

The theorem states that if \lstinline{approxAsPoly} succeeds and returns
$q_{f,n}$ and $\delta_{f,n}$, then the approximation error on input range $I$
between $f(x)$ and $q_{f,n} (x)$ is upper bounded by $\delta_{f,n}$.

In the rest of this section, we first explain how Dandelion automatically
computes truncated Taylor series for elementary functions like $\Sin$ and
$\Exp$, then we explain how Dandelion extends this approach in
\lstinline{approxAsPoly} to compute a single polynomial approximations of
more complicated elementary function expressions like $\Exp(x * \frac{1}{2})$
via interval analysis and
propagation of polynomial errors.
Throughout this section, we use $f$ to refer to the elementary function from
the certificate, $t_{f,n}$ as truncated Taylor series, $\delta_{t,n}$ as the
approximation error of the series, $q_{f,n}$ as polynomial approximation, and
$\delta_{f,n}$ as the overall approximation error.

\subsection{Truncated Taylor Series for Single Elementary Functions}

Both $t_{f,n}$ and $\delta_{t,n}$ depend on the approximated elementary
function $f$, as well as the number of series terms $n$ from the
certificate.
Overall, Dandelion automatically computes a truncated Taylor series for
the elementary functions \Sin{}, \Cos{}, \Exp{}, \Atn{}, and \Log{}\footnote{Dandelion currently does not support $tan$, as a straight-forward reduction to \Sin{}(x)/\Cos{}(x) did not work out. We plan to incorporate a more direct series from HOL-Light in the future.}; the series
expansions for \Exp{} and \Log{} already existed in HOL4 prior to Dandelion and
we port the series for \Atn{} from HOL-Light.
For \Sin{} and \Cos{} we prove series based on textbook descriptions.
Formally, Dandelion proves a truncated Taylor series for each elementary
function once and for all as
\begin{thm}\label{thm:truncatedTaylor}
$\forall x\;n.\;Pre (x) \Rightarrow f (x) = \sum_{i=0}^{n}
  (\frac{f^{i}\;0}{i!} * x^{i}) + \delta_{t, n}(x)$
\end{thm}
Here, $f^{i}$ is the $i$-th derivative of $f$, and the approximation
error $\delta_{f,n}(x)$ is soundly bounded from the remainder term of Taylor's
theorem for input value $x$.
Predicate $Pre$ is a precondition constraining the interval on which function
$f$ can be approximated by the truncated series.
When approximating an elementary function $f$ by its truncated series, Dandelion
always ensures that this precondition $Pre$ is true:
The series for \Exp{} requires inputs to be non-negative, the series for \Log{}
requires arguments greater than $1$, and \Atn{} requires arguments in $(-1,1)$.
The series for \Sin{} and \Cos{} have no preconditions.

At certificate checking time Dandelion automatically computes an upper bound to
the approximation error $\delta_{t,n}(x)$.
Further, the second phase of Dandelion operates on polynomials following
Harrison's formalization.
Therefore, we prove once and for all that the truncated series from
\autoref{thm:truncatedTaylor} can be implemented in Harrison's polynomial
datatype:

\begin{thm}\label{thm:taylorSum}
$\forall n. \sum_{i=0}^{n} (\frac{f^{i}\;0}{i!} * x^{i}) = t_{f,n}\;(x)$
\end{thm}

\autoref{thm:taylorSum} proves that $t_{n}$ implements the truncated Taylor
series on the left-hand side for an arbitrary number of approximation steps $n$.
We prove versions of \autoref{thm:taylorSum} for each elementary function
supported by Dandelion.
Finally, the proof of First Phase Soundness (\autoref{thm:approxPolySound})
for a single elementary function is a simple combination of
Theorems~\ref{thm:truncatedTaylor} and~\ref{thm:taylorSum}.

\subsection{Approximations of More Complicated Expressions}

Next, we explain how Dandelion uses truncated Taylor series to approximate
more complicated elementary function expressions, using $\Exp(y \times \frac{1}{2})-1$
on the interval $[1,2]$ as an example\footnote{Currently, Dandelion does not
generally support divisions, hence we represent $\frac{y}{2}$ as $y \times \frac{1}{2}$ explicitly.}.
In general, a Remez-like algorithm can return an approximation for a compound
function or an expression, as long as it stays univariate.
Compared to approximating individual functions, \eg{} \Exp{}, an overall
expression approximation can be more accurate, sometimes avoiding undesirable
effects such as cancellation.
Hence, Dandelion should also be able to certify those.

From Theorems~\ref{thm:truncatedTaylor} and~\ref{thm:taylorSum} Dandelion
knows how to automatically compute a polynomial approximation $t_{\Exp{}, n}$
and an approximation error $\delta_{\Exp,n}(x)$ for the exponential function
for a given input range on the argument.
In our example, the input argument is $y \times \frac{1}{2}$, and thus the value of
$\delta_{\Exp, n}(x)$ depends on the range of this expression, which Dandelion
computes automatically using interval arithmetic~\cite{intervals}.

As interval analysis, we reuse an existing HOL4 formalization~\cite{becker2018verified},
and extend it with range bounds for elementary functions.
For our example Dandelion also needs to compute a range bound for
$\Exp{}(y \times \frac{1}{2})$.
In general, because elementary functions are defined non-computably in HOL4, we
have to rely on a trick to compute interval bounds.
To compute interval bounds for elementary functions, Dandelion reuses our
formalized truncated Taylor series.
From \autoref{thm:truncatedTaylor} and \autoref{thm:taylorSum}, we derive for
$f$ that
\begin{equation}
| f(x) - t_{t,n} (x) | \leq \delta_{t,n}
\end{equation}
From this inequality, we derive a bound on $f(x)$ in the interval $[a,b]$
\begin{equation}\label{eqn:ivbounds}
  t_{t,n} (a) \leq f(x) \leq t_{t,n} (b) + \delta_{t,n}
\end{equation}
\autoref{eqn:ivbounds} holds for monotone $f$ only, and thus we cannot apply it
to $\Sin$ and $\Cos$ as they are periodic.
For both functions, interval analysis returns the closed interval $[-1,1]$.
Dandelion's interval analysis is proven sound once and for all in HOL4.

With the interval analysis, we can soundly compute a polynomial approximation
for $\Exp$, $t_{\Exp,n}$ on the range of $y \times \frac{1}{2}$.
Dandelion automatically composes the polynomial $y \times \frac{1}{2}$
with $t_{\Exp,n}$ to obtain a polynomial $q_{\Exp(y \times \frac{1}{2}),n}$ with
approximation error $\delta_{\exp,n}(x)$.
However, we still need to come up with a polynomial approximation $p$ and an
approximation error for the full function $\Exp(y \times \frac{1}{2}) - 1$.
In our example, Dandelion treats the constant $1$ as a polynomial returning
$1$, and automatically computes the polynomial difference of
$q_{\Exp(\ldots), n}$ and $q_{1,n}$.
The global approximation error $\delta$ for the difference of $q_{\Exp(\ldots),n}$
and $q_{1,n}$ depends on the approximation errors accumulated in both
polynomials.
In a final step, Dandelion automatically computes an upper bound on the
global approximation error by propagating accumulated errors through the subtraction
operation.

Generally, Dandelion implements an automatic approximation error analysis inside
function \lstinline{approxAsPoly} that propagates accumulated approximation
errors.
The propagation is implemented for basic arithmetic, and elementary
functions to support \eg{} expressions like $\Exp(x) + \Sin (x - 1)$.

\paragraph*{Computing Propagation Errors for Sin and Cos}
To accurately propagate approximation errors through \Sin{} and \Cos{},
our soundness proof assumes that the accumulated approximation error is
contained in the interval $[0, \frac{\pi}{2}]$.
This does not pose a true limitation of Dandelion as errors larger than $\pi/2$ would
anyway be undesirable and impractical.
For the correctness proof of \lstinline{approxAsPoly} (\autoref{thm:approxPolySound}), however, Dandelion must
automatically prove that accumulated errors are less than or equal to $\pi/2$.
This poses a challenge as in HOL4 $\pi$ is defined non-computably using
Hilbert's choice operator:
if $0 \leq x \leq 2$ and $\Cos(x) = 0$, then $\pi$ is $2 \times x$.
To solve this problem, we reuse the truncated Taylor series of $\Atn$
and the fact that $\Atn (1) = \pi / 4$ to compute a lower bound $r$ in HOL4,
where $r \leq \pi$.
At certificate checking time, when propagating the error $\delta_{f,n}$ through
\Sin{} and \Cos{}, Dandelion checks $\delta_{f,n} \leq \frac{r}{2}$, which by transitivity
proves that $\delta_{f,n} \leq \frac{\pi}{2}$.

\subsection{Extending Dandelion's First Phase}

All of the truncated Taylor series proven in Dandelion are for single applications
of an elementary function.
For a particular application it may be beneficial to add special cases
to compute a single, more accurate, truncated Taylor series of an elementary
function like $\Exp (\Sin (x))$ instead of computing a truncated series for each
function separately.

In Harrison's original proof this would require manually redoing a large
chunk of the proof work whereas for Dandelion such an extension amounts to 4
steps:
Proving the truncated Taylor series as in \autoref{thm:truncatedTaylor},
implementing and proving correct the polynomial $t_{f,n}$ as in
\autoref{thm:taylorSum}, extending \lstinline{approxAsPoly} with the special case
for the new elementary function, and finally
using the theorems proven for the first two steps to extend First Phase Soundness
(\autoref{thm:approxPolySound}) with a correctness proof for the new case.
Complexity of the proofs only depends on the complexity of the series
approximation.
Dandelion then automatically uses the new series approximation whenever the
approximated function is encountered in a certificate,
and the global soundness result of Dandelion still holds without any required
changes.
The second phase directly benefits from adding additional approximations as
more accurate Taylor series decrease the approximation error of the first phase.
%%% Local Variables:
%%% mode: latex
%%% TeX-master: "../main"
%%% End:

%% file: sections/sturmSeq.tex
% !TEX root = ../main.tex
\section{Validating Polynomial Errors}\label{sec:sturmSeq}
For a certificate consisting of an elementary function \lstinline{f},
polynomial approximation \lstinline{p}, approximation error $\varepsilon$,
input constraints \lstinline{I}, and truncation steps \lstinline{n},
the first phase of Dandelion computes a truncated Taylor series $q_{f,n}$ and an
approximation error $\delta_{f,n}$, which is sound by
\autoref{thm:approxPolySound}.
Both $q_{f,n}$ and \lstinline{p} are polynomials, and following Harrison's
terminology, we refer to their difference $q_{f,n} (x) - \texttt{p} (x)$
as the error polynomial $h(x)$.
In the second phase, Dandelion automatically finds an upper bound to the
extremal values of $h(x)$ and compares this upper bound to the residual
approximation error $\varepsilon - \delta_{f,n}$, which we refer to as $\gamma$.
We prove soundness of the second phase once and for all as a HOL4 theorem:
\begin{thm}[Second Phase Soundness]\label{thm:errPoly}
  \begin{align*}
    \forall x.\;x \in \texttt{I}(x) \Rightarrow |\,q_{f,n} (x) - \texttt{p} (x)\,| \leq \gamma
  \end{align*}
\end{thm}

Before going into the details of how Dandelion automatically validates
the residual error $\gamma$, we quickly recall the key real analysis
result which we rely on in this phase:
on a closed interval $[a,b]$, a
differentiable polynomial $p$ can reach its extremal values at the outer points
$p(a)$, $p(b)$, and the zeros of $p$'s first derivative $p'$ (\autoref{thm:fstderiv}).
To find the extremal values of $h(x)$, Dandelion thus needs to automatically find
\emph{all} zeros of $h'(x)$.

Dandelion splits finding the extremal values and validating $\gamma$ into three
automated steps:
\begin{enumerate}
\item Compute the number of zeros using Sturm's theorem (\autoref{thm:sturm} in \autoref{sec:overview})
\item Validate a guess of the zeros computed by an unverified, external oracle
\item Compute an upper bound on extremal values (using the validated zeros) and compare with $\gamma$
\end{enumerate}
Conceptually, the second phase automates the main part of Harrison's manual
proof, and the key step is computing Sturm sequences automatically in the first
step.
Next, we explain the ideas behind automating each of the steps.

\subsection{Bounding the Number of Zeros of a Polynomial}

Dandelion bounds the number of zeros of a polynomial
using Sturm's theorem (\autoref{thm:sturm}).
A key challenge in developing this part of Dandelion was ensuring that the Sturm
sequence is computable inside HOL4.
In HOL-Light, Harrison defines Sturm sequences as a non-computable predicate
\lstinline{STURM} that existentially quantifies results, and thus can only be
used to validate results in a manual proof.

\begin{figure}
\begin{lstlisting}[escapechar=\%,basicstyle=\footnotesize\ttfamily,numbers=left,numberstyle=\tiny]
sturm_seq (p, q, n) =
  if n = 0 then
    if (rm (p, (%$\frac{1}{q[\texttt{deg} q]}$%) * q) = 0 %$\wedge$% q <> 0) then SOME []
    else None
  else let g = - (rm (p, (%$\frac{1}{q[\texttt{deg} q]}$%) * q)) in
   if g = 0 %$\wedge$% ~ q = 0 then Some []
   else if (q = 0 %$\vee$% (deg q < 3)) then None
   else case sturm_seq (q, g, n-1) of
     None => None
     |Some ss => Some (g::ss)
\end{lstlisting}
\caption{HOL4 definition of Sturm sequence computation}\label{fig:sturmSeqComp}
\end{figure}

In \autoref{fig:sturmSeqComp}, we show how Dandelion computes Sturm sequences.
Function \lstinline{rm (p,q)} computes the remainder of the polynomial division
of \lstinline{p} by \lstinline{q}, \lstinline{deg p} is the degree of polynomial
\lstinline{p}, and \lstinline{q[n]} is the extraction of the \lstinline{n}-th
coefficient of \lstinline{q}.
As each polynomial division operation decreases the degree of the result by at
least 1, the Sturm sequence for a polynomial $p$ has a maximum length of
$\text{deg}(p)-1$, as computation starts with $p$ and its first derivative $p'$.
Function \lstinline{sturm_seq} is therefore initially run on polynomial
$h'(x)$, $h''(x)$, and $\text{deg}(h') - 1$.

\begin{sloppypar}
If \lstinline{sturm_seq(deg h'-1, h', h'')} returns list \lstinline{sseq},
the complete Sturm sequence is \lstinline{h'::h''::sseq}, and Dandelion
computes the number of zeros of $e'$ as its variation on the input range,
based on \autoref{thm:sturm}.
\end{sloppypar}

We have proven once and for all that the results obtained from
\lstinline{sturm_seq(n, p', p'')} satisfy Harrison's non-computable predicate \lstinline{STURM}.
Thus we can reuse Harrison's proof of Sturm's theorem (\autoref{thm:sturm}).
Harrison's Sturm sequences also use on a non-computable predicate for defining
the result of polynomial division, and we prove it equivalent to a computable
version in Dandelion, inspired by the one provided by Isabelle/HOL~\cite{isabelle}.
Ultimately, Dandelion uses these two equivalence proofs to reuse Harrison's
proof of Sturm's theorem which we ported from HOL-Light.

\subsection{Finding Zeros of Polynomials}

Given the numbers of zeros $nz$ for $h'(x)$, Dandelion next finds their values.
As zero-finding is highly complicated even in non-verified settings, Dandelion
uses an external oracle to come up with an initial guess of the zeros.
These initial guesses are presented as a list of confidence intervals $[a,b]$,
where $h'(x)$ has a zero between $a$ and $b$.
Further, the algorithm computing the confidence intervals need not be verified,
as the result can easily be validated by Dandelion.
To validate a list of guesses $Z$, Dandelion again relies on a result from
real-number analysis, proven by Harrison:
For a confidence interval $[a,b]$, function $f$ has a zero in the interval, if
its first derivative $f'$ changes sign in the interval $[a,b]$.

Dandelion validates the confidence intervals $Z$ using a computable function
that checks automatically for each element $[a,b]$ in $Z$, that
$h''(a) \times h''(b) \leq 0$, which is equivalent to a sign-change in the
interval.
If the number of zeros found is $nz$, Dandelion checks that this sign change
occurs at least $nz$ times in $Z$.

While we do prove our approach for finding zeros of polynomials sound,
Dandelion is necessarily incomplete.
One known source for incompleteness are so-called multiple roots as they occur
\eg{} in $p (x) = (x - 1)^2$.
Harrison's formalization implicitly relies on the polynomial being squarefree
and Dandelion inherits this limitation.
This issue could potentially be addressed using the approach of Li
\etal{}~\cite{li2019deciding} though it would require reproving Sturm's theorem
for non-squarefree polynomials.

\subsection{Computing Extremal Values}

In the final step, Dandelion uses the validated confidence intervals $Z$ which
contain all zeros of $h'(x)$ to compute an upper bound to the extremal values of
$h(x)$.
For interval $[a,b]$, Dandelion would ideally bound the error of $h(x)$ in $[a,b]$
as the maximum of $h(a)$, $h(b)$, and $h(y)$, where $y$ is a zero of $h'(x)$.
However, we have only confidence intervals for the zeros, and not their exact
values available.
Therefore, Dandelion's computation of an upper bound to $e(x)$ is more involved,
and we base it on a theorem of Harrison.
Harrison's theorem is a generalization of \autoref{thm:fstderiv} for polynomial $p$,
with derivative $p'$, on interval $[a,b]$:
\begin{thm}\label{thm:approxZeros}
  \begin{align*}
  (1) \quad{} &(\forall x.\; a \leq x \leq b \wedge f' (x) = 0 \Rightarrow
                \exists (u,v).\;(u,v) \in Z \wedge u \leq x \leq v)\;\wedge\\
  (2) \quad{} &(\forall x.\;a \leq x \leq b \Rightarrow |\,p'(x)\,| \leq B)\;\wedge\\
  (3) \quad{} &(\forall [u,v].\;[u,v] \in Z \Rightarrow
                a \leq u \wedge v \leq b \wedge |\,u - v\,| \leq e \wedge |\,f (u)\,| \leq K)\;\Rightarrow\\
    &\forall x.\;a \leq x \leq b \Rightarrow |\,p (x)\,| \leq \max (|\,f(a)\,|, |\,f(b)\,|, K + B \times e)
  \end{align*}
\end{thm}

The theorem can be used to prove an upper bound on the error polynomial $h(x)$
which then can be compared to the residual error $\gamma$.
For Dandelion, we automatically computed the values described by the assumptions
to compute an overall bound on $h(x)$.
We implement this computations in a function \lstinline{validateErr}, and
explain each of its computation steps on a high-level, based on the assumptions
of \autoref{thm:approxZeros}.

The first assumption $(1)$ from \autoref{thm:approxZeros} states that the
confidence intervals in $Z$ contain only valid zeros and has been established
automatically by the previous step.
Based on assumption $(2)$, Dandelion computes $B$ by evaluating $|h'(x)|$ on
$\max (|a|, |b|)$.
Following assumption $(3)$, Dandelion computes $K$ as the maximum value of
evaluating the error polynomial $h$ on the lower bounds of the confidence
intervals in $Z$, and a value $e$ as the maximum value of $|u - v|$ for each
$[u,v]$ in $Z$.
Dandelion then computes the overall bound on the error polynomial $h$ as
$\max (h(a),\,h(b),\,K + B \times e)$.
To validate the residual error $\gamma$, it then suffices to check
$\max (h(a),\,h(b),\,K + B \times e) \leq \gamma$.

Overall, we prove once and for all soundness of Dandelion as
\begin{thm}[Dandelion Soundness]\label{thm:Dandelion_corr}
  \begin{align*}
    {\normalfont \texttt{Dandelion(f,p,I,}}\varepsilon{\normalfont \texttt{,n)}} = {\normalfont \texttt{true}}\;\Rightarrow\;
    (\forall x.\;x \in {\normalfont \texttt{I}} \Rightarrow | {\normalfont \texttt{f}} (x) - {\normalfont \texttt{p}} (x) | \leq \varepsilon)
  \end{align*}
\end{thm}

The proof of \autoref{thm:Dandelion_corr} uses the triangle inequality to
combine the theorem First Phase Soundness (\autoref{thm:approxPolySound})
with the theorem Second Phase Soundness (\autoref{thm:errPoly}).

%\begin{figure}
%\begin{lstlisting}[escapechar=?, numbers=left]
%validateZerosLeqErr e I numZeros zeros eps =
%  let mAbs = max (abs (fst I)) (abs (snd I));
%    realZeros = findN numZeros
%      (?$\lambda$? (u,v). poly (diff e) u * poly (diff e) v ?$\leq$? 0) zeros;
%    B = poly (MAP abs (diff e)) mAbs;
%    e = getMaxWidth realZeros;
%    ub = getMaxAbsLb e realZeros;
%    globalErr = max (abs (poly e (fst I)))
%                    (max (abs (poly e (snd I)))
%                     (ub + B * e))
%  in if ?$\neg$? (validBounds I realZeros ?$\wedge$? recordered (fst I) realZeros (snd I)) then
%        (Invalid "Zeros not correctly spaced", 0)
%    else if LENGTH realZeros < numZeros then
%        (Invalid "Did not find sufficient zeros", 0)
%    else if globalErr ?$\leq$? eps
%    then (Valid, globalErr)
%    else (Invalid "Bounding error too large", 0)
%\end{lstlisting}
%\caption{Function for computing a bound on the error polynomial $e(x)$ in Dandelion}\label{fig:validateZerosLeqErr}
%\end{figure}
%%% Local Variables:
%%% mode: latex
%%% TeX-master: "../main"
%%% End:

%% file: sections/implementation.tex
% !TEX root = ../main.tex
\section{Extracting a Verified Binary with CakeML}\label{sec:implementation}
Computations performed in interactive theorem provers are known to be slower
than those in unverified languages.
To alleviate this performance problem, the proof-producing synthesis~\cite{JAR20}
implemented in the CakeML verified compiler~\cite{CakeML} translates HOL4
functions into their CakeML counterpart, with an equivalence proof.
These translated CakeML functions are compiled into machine code with
the CakeML compiler, and as CakeML is fully verified, the machine code enjoys
the same correctness guarantees as its HOL4 version.

During an initial test run we noticed that the Sturm sequence computations in
the second phase are the most computationally expensive task of Dandelion.
HOL4 represents real-numbers as (reduced) fractions during computation, and we
noticed that in Dandelion their size still grew quite large, leading to a single
multiplication taking up to 6 hours.
%The second phase of Dandelion is the most computation intensive, especially so
%for the computation of Sturm sequences.
%During an initial test run using plain HOL4 computations, we found that a single
%real-number multiplication could take up to 6 hours.
%This is due to HOL4 representing real-numbers as fractions during computation, and
%even though they are reduced, we noticed that their representation size grew
%quickly in Dandelion.
% \ed{explain why? because they need to be rationals?}
% \hb{Good enough explanation?}
Therefore, we use CakeML's proof-producing synthesis to extract a verified
binary for the computations described in \autoref{sec:sturmSeq}.
To communicate results of the first phase with the binary, we implemented an
(albeit unverified) lexer and parser that reads-in results from the first phase.

%Most of the development time for the second phase was spend defining computable
%versions of polynomial division and Sturm sequences, and proving them equivalent
%to their non-computable counterparts defined by Harrison.
%While we see the appeal of defining non-computable, existentially quantified
%versions of polynomial divisions and Sturm sequences, if there had been a
%computable version from the get go, these versions may have simplified some of
%the manual proofs that Harrison performed, and in general we argue that
%computable functions with a correctness proof should be favoured over
%non-computable predicates to make developments easier to reuse in the long run.
%\ed{Is it possible to rephrase the above a little to make sounds a bit less - judgemental?
%I can imagine that this may upset some people, since it sounds a bit like a philosophical/personal
%perspective.}
%%% Local Variables:
%%% mode: latex
%%% TeX-master: "../main"
%%% End:

%% file: sections/evaluation.tex
% !TEX root = ../main.tex
\section{Evaluation}\label{sec:evaluation}

We have described how Dandelion automatically validates polynomial
approximations from Remez-like algorithms.
Next, we demonstrate Dandelion's usefulness with three separate experiments, by
demonstrating that Dandelion fully automatically
\begin{enumerate}
\item validates certificates generated with an off-the-shelf Remez-like algorithm (\autoref{subsec:backend_verif})
\item validates certificates for more complicated elementary function expressions (\autoref{subsec:cmpound_verif})
\item validates certificates for less-accurate techniques (\autoref{subsec:chebyshev_verif})
\end{enumerate}
All the results we report in this section where gathered on a machine running
Ubuntu 20.04, with an 2.7GHz i7 core and 16 GB of RAM.
All runnning times are measured using the UNIX \lstinline{time} command as
elapsed wall-clock time in seconds.

\subsection{Validating Certificates of a Remez-like Algorithm}\label{subsec:backend_verif}
In our first evaluation, we show that Dandelion certifies accurate approximation
errors from an off-the-shelf Remez-like algorithm.
We generate certificates by combining the Daisy static analyzer with the
Sollya approximation tool~\cite{Sollya}, and extend Daisy with a simple pass
that replaces calls to elementary functions with an approximation computed by
Sollya.
As a Remez-like algorithm, we use the \lstinline{fpminimax}~\cite{brisebarre2007efficient}
function in Sollya.
Our pipeline is benchmarked on numerical kernels taken from the FPBench
benchmark suite~\cite{fpbench}, and the benchmarks used by an unverified
extension of Daisy with approximations for elementary
functions~\cite{izycheva2019synthesizing}.
These benchmarks represent kernels as they occur in \eg{} embedded systems,
and thus they benefit from custom polynomial approximations.
The original work by Izycheva \etal{}~\cite{izycheva2019synthesizing}
synthesizes polynomial approximations whose
target error bounds are usually larger than those inferred by Sollya. Hence,
Dandelion could validate Daisy's bounds as well, but for the sake of the evaluation
we choose more challenging, tighter bounds.
For each benchmark, Daisy creates a certificate for each approximated elementary
function, amounting to a total of 96 generated certificates.

Sollya's implementation of \lstinline{fpminimax} can be configured to use different
degrees for the generated approximation and different formats for the coefficients
of the approximation.
In our evaluation we approximate elementary functions with a degree 5 polynomial,
storing the coefficients with a precision of 53 bits.
All input ranges used in the certificates are computed by Daisy without
modifying them, except for $\Exp$, where we disallow negative intervals, \ie{}
if Daisy wanted to approximate on $[-x, y]$, we change it to $[0, y]$ as
Dandelion currently does not support negative exponentials.
This can be fixed by straight-forward range reductions that are independent to
the approximations computed by Remez-like algorithms.
For each such approximation, Daisy creates a certificate to be checked by
Dandelion.

The MetiTarski automated theorem prover~\cite{akbarpour2010metitarski} is a tool
that provides the same level of automation as Dandelion, but relies on a
different technology for proving inequalities.
MetiTarski is based on the Metis theorem prover~\cite{hurd2003d}
and can output proofs in TSTP format~\cite{sutcliffe2004tstp}.
It relies on an external decision procedure to discharge some goals,
and for our experiments we used Mathematica.
In general, MetiTarski checks real-number inequalities that may contain
elementary functions, thus we compare the number of certificates validated
by Dandelion to those that can be checked by MetiTarski.

Further, the CoqInterval package~\cite{martin2016coqinterval} proves
inequalities about elementary functions in the Coq theorem prover~\cite{coq}.
In contrast to Dandelion's technique based on high-accuracy Taylor polynomials
and Sturm sequences, CoqInterval is based on interval arithmetic with optional
interval bisections and high-accuracy Taylor polynomials.
Further, Dandelion is verified in HOL4, while CoqInterval is implemented in Coq.
To compare the two approaches, our evaluation also includes the CoqInterval
package.
Our evaluation excludes a similar approach formalized in Isabelle/HOL~\cite{holzl2009proving}
because we could not come up with a straight-forward translation of our
certificates as inputs to the tool.
We have manually tested some of our benchmarks and expect it to produce results
similar to CoqInterval.

\begin{table}
  \centering
\begin{tabular}{lrrrrrrrr}
  \toprule
 Function & \# & \multicolumn{3}{c}{Dandelion} & \multicolumn{2}{c}{MetiTarski} & \multicolumn{2}{c}{CoqInterval}\\
& & Verified & HOL4(s) & Binary(s) & Verified & Time(s) & Verified & Time(s)\\
  \midrule
atan & 2 & 1 & 11.62 & 20.36 & 2 & 6.80 & 2 & 1.74 \\
cos & 28 & 25 & 202.35 & 251.33 & 25 & 3.15 & 26 & 1.91 \\
exp & 21 & 18 & 39.58 & 212.54 & 10 & 5.35 & 20 & 1.58 \\
log & 8 & 0 & 0 & 0 & 5 & 4.99 & 8 & 1.68 \\
sin & 31 & 27 & 17.83 & 295.66 & 25 & 6.32 & 31 & 1.78 \\
  \midrule
Total & 90 & 71 & & & 67 & & 87 &\\
%sqrt & 3 & 0 & 0 & 0 & 0 & 0 \\
%tan & 3 & 0 & 0 & 0 & 0 & 0 \\
  \bottomrule
\end{tabular}
\caption{Overview of certificates validated with Dandelion, MetiTarski, and CoqInterval}\label{tbl:exp}
\end{table}

Our results are given in \autoref{tbl:exp}.
The left-most column of \autoref{tbl:exp}, contains the name of the elementary
function approximated by Daisy, and the second column, labeled with a \# contains the
number of certificates generated for the elementary function, with unique input
ranges.
The next three columns, headed ``Dandelion'', contain the number of certificates
validated by Dandelion, the average HOL4 running time for the first phase in seconds,
and the average running time of the binary for the second phase in seconds.
The next two columns, headed ``MetiTarski'', contain the number of certificates
validated by MetiTarski, and the average running time in seconds.
The final two columns, headed ``CoqInterval'', contain the number of certificates
validated by CoqInterval, and the average running time in seconds.

Our evaluation truncates Taylor series after $32$ terms in \lstinline{approxAsPoly}.
In general, we found six times the degree of the computed approximation to be a
good estimate for when to truncate Taylor series in Dandelion.
Our use of $32$ instead of $30$ is a technical detail, as some Taylor series
require both the number of series terms $n$, as well as $\frac{n}{2}$ to be even.
In general, the number of series terms has to be significantly higher than the
degree of the approximated polynomial, to make the approximation error of the
first phase almost negligible.

Overall, we notice that each of the tools in our evaluation certifies a slightly different
set of approximations.
In total, CoqInterval certifies most of the benchmarks, but Dandelion
successfully checks one cosine certificate that CoqInterval fails to check.
While Dandelion validates more certificates than MetiTarski, both MetiTarski and
CoqInterval validate certificates that are currently out of reach for Dandelion.
This is mostly due to the first phase of Dandelion.
Even though we used general, widely known truncated Taylor series for
all supported elementary functions, Dandelion fails to validate certificates
for the $\Log$ function.
We have inspected the generated certificates, and Dandelion cannot compute a
high-accuracy polynomial approximation for 5 of them because they do not satisfy
the precondition of Dandelion's Taylor series.
For the remaining 3 certificates, we ran into issues with Sollya's computation
of the confidence intervals for the zeros.
On a high-level, the problem originates from the derivative of the error
polynomial $h(x)$ being very close to 0, leading to a huge number of zeros
being found, \ie{} computation not terminating within a reasonable amount of
time.
To demonstrate that Dandelion still certifies errors for the $\Log$ function, we
add an example in \autoref{subsec:cmpound_verif}.

While Dandelion cannot certify errors for the $\Log$ function in this part of
the evalution, this is not a conceptual limitation, as polynomial approximations
are commonly paired with an argument reduction strategy.
While verification of these strategies is orthogonal to validating results
of an Remez-like algorithm, they could be used to reduce the input range of the
approximated elementary function into a range that Dandelion can certify.
More generally, we have done the heavy lifting of automating the computations
and implementing the general framework, such that adding more accurate Taylor
series to Dandelion amounts to mere proof engineering, modulo coming up with
Taylor series in the first place.
For the certificates for $\Atn$, $\Cos$, $\Sin$, and $\Exp$, we notice that the
average running time is in the order of minutes, making certificate checking
with Dandelion's verified binary feasible.

We compare the approximation errors recorded in the certificates with the
infinity norm computed by Sollya, which is the most-accurate estimate of the
approximation error~\cite{chevillard2011}.
Overall, Dandelion certifies an approximation error in the same order of
magnitude as the infinity norm for 61 certificates.
For the remaining 10, the error is a sound upper bound.
In general, infinity norm-based estimates are known to be the most accurate and
their verification requires more elaborate techniques than Sturm
sequences~\cite{chevillard2011}.
Consequently we would not expect Dandelion to be able to always certify infinity norms.

We ran the evaluation for an approximation degree of $3$, with precisions of $53$ and
$23$ each to measure the influence of those parameters.
Overall, the running time significantly decreases when decreasing from degree 5 to 3,
going from average running times of minutes to average running times of seconds.
Decreasing the precision of the coefficients further speeds up evaluation,
though not as significant as decreasing the degree did.
This suggests that higher coefficient accuracies can easily be used for generating
polynomials with Remez-like algorithms, and lower degree polynomials should be
preferred for fast validation.

\subsection{Validating Certificates for Elementary Function Expressions}\label{subsec:cmpound_verif}

Next, we show that Dandelion can also certify approximation errors for complicated
elementary function expressions.
We validate with Dandelion approximation errors for random examples involving
elementary functions and arithmetic.
Polynomial approximations are again generated by Sollya.

\begin{table}
  \centering
  \begin{tabular}{l r r r r r r r}
    \toprule
    Function & Range & Deg. & Prec. & $\infty$-norm & Error & HOL4 & Binary\\
    \midrule
    $\cos(x + 1)$ & $[0, 2.14]$ & 5 & 53 & 3.06E-5 & 3.06E-5 & 169 & 63 \\
    $\sin(x - 2)$ & $[-1, 3.00]$ & 5 & 53 & 2.05E-3 & 2.91E-3 & 93 & 68 \\
    $\ln(x + \frac{1}{10})$ & $[1.001, 1.1]$ & 3 & 32 & 1.08E-7 & 1.08E-7 & 2775 & 1773\\
    $\exp(x \times \frac{1}{2}) + \cos(x \times \frac{1}{2})$ & $[0.1, 1.00]$ & 5 & 53 & 2.03E-9 & 4.45E-9 & 711 & 5 \\
    $\arctan(x) - \cos (\frac{3}{4} \times x)$ & $[-0.5, 0.5]$ & 5 & 53 & 1.18E-5 & 1.18E-5 & 24 & 2308 \\
    \bottomrule
  \end{tabular}
  \caption{Functions approximated with Sollya using \lstinline{fpminimax} and certified with Dandelion}\label{tbl:compRes}
\end{table}

An overview of our results is given in \autoref{tbl:compRes}.
The table shows the approximatied function, then Sollya's parameters (the input range,
the target degree (Deg.), the target precision (Prec.)), and then the
infinity-norm ($\infty$-norm) of the approximation.
The final columns summarize the Dandelion results, giving the certified
approximation error, and the running time in seconds of the first phase (HOL4) and the
second phase (Binary).

Overall, we notice that the certified approximation error is on the same order
of magnitude as the infinity norm for all examples.
We also notice that performance for both phases varies across the different
examples.
For the first phase this is often due to how the input ranges are encoded in the
certificate.
We noticed that HOL4 is very sensitive to how the fractions representing
real numbers are encoded when performing computations.
Similarly, performance of the second phase greatly varies depending on the
complexity of the error polynomial computed by the first phase.
We observe that perfomance improves with both smaller degree polynomials, and
smaller representations of the polynomial coefficients.

The results in \autoref{tbl:compRes} exclude examples where two elementary functions
are composed with each other, \eg{} as in $\Exp(\Sin (x - 1))$.
This is because Dandelion computes the global high-accuracy approximation in the first phase
via polynomial composition of an approximation for $\Exp$ and $\Sin$.
While this is theoretically supported by Dandelion, we found that the polynomial
composition leads to an exponential blow-up in the degree of the error polynomial.
Even for the innocuously looking example $\Exp(\Sin (x-1))$, the second phase
could not validate a polynomial approximation within 24 hours.
This clearly motivates the use of more elaborate Taylor series if compound
elementary functions need to be certified by Dandelion.
In general, settings where elementary functions like those in \autoref{tbl:compRes}
are used could potentially be made more accurate and be validated faster with
custom Taylor series.

\subsection{Validating Certificates for Simpler Approximation Algorithms}\label{subsec:chebyshev_verif}
Remez-like algorithms are known to be the most accurate approximation algorithms.
However, less accurate approaches are still in use today, and as such
interesting targets for verification.
Br{\'e}hard \etal~\cite{brehard2019certificate} certify Chebyshev
approximations in the Coq theorem prover, where their approach requires some
manual proofs.
We demonstrate that Dandelion also certifies Chebyshev approximations on some
random examples by computing Chebyshev approximations with Sollya's function
\lstinline{chebyshevform}.
The results are shown in \autoref{tbl:chebyshev}.

Again, we first give Sollya's parameter and the infinity norm, then we give
the error certified by Dandelion, and the execution times for the first and
second phase.

We also include the approximation certified by Harrison~\cite{harrison1997verifying},
labeled with a $^{*}$.
The polynomial has degree 3, but we leave the precision empty as it is not
generated by Sollya, and we do not provide an infinity norm.
The only difference to the proof from Harrison is that we prove the bound only
for positive $x$, as Dandelion currently does not handle exponentials on
negative values.
The lower bound of the range is $0.003$, instead of $0$ to rule out a
$0$ on the lower bound (as $\exp(x) - 1 = 0$ for $x = 0$), which we must exclude
by \autoref{thm:approxZeros}.
Harrison's manual proof of the polynomial approximation then reduces to a
single line running Dandelion on the encoding.

\begin{table}[t]
  \centering
  \begin{tabular}{l r r r r r r r}
    \toprule
    Function & Range & Deg. & Prec. & $\infty$-norm & Error & HOL4 & Binary\\
    \midrule
    $\cos(x)$ & $[0, 2.14]$ & 5 & 53 & 3.17E-7 & 3.22E-7 & 169 & 63 \\
    $\sin(x + 2)$ & $[-1.5, 1.5]$ & 5 & 53 & 4.47E-4 & 7.60E-4 & 142 & 138 \\
    $\sin(3 \times x) + \exp (x \times \frac{1}{2})$& $[0,1]$ & 3 & 53 & 2.45E-2 & 2.48E-2 & 54 & 1897 \\
    $\exp(x) - 1^{*}$ & $[0.003, 0.01]$ & 3 & & & $2^{-33.2}$ & 133 & <1 \\
    \bottomrule
  \end{tabular}
  \caption{Chebyshev approximations certified with Dandelion}\label{tbl:chebyshev}
\end{table}
%%% Local Variables:
%%% mode: latex
%%% TeX-master: "../main"
%%% End:

%% file: sections/related.tex
% !TEX root = ../main.tex
\section{Related Work}

Throughout the paper, we have already hinted at the immediate related work.
Next, we explain the key conceptual differences between Dandelion and the
immediate related work and put Dandelion into the greater context.
In general, Dandelion touches upon two key research areas in interactive and
automated theorem proving:
techniques for approximating elementary functions and techniques for proving
theorems involving real-numbered functions.

\emph{Approximating Elementary Functions}
The work on approximating elementary functions can be distinguished among two
axes: whether or not the work provides rigorous machine-checked proofs, and
whether the work is fully automated or requires user intervention.

Fully automated, rigorous machine-checked proofs, similar to Dandelion
are provided by the work by Br{\'e}hard \etal{}~\cite{brehard2019certificate}.
They develop a framework for proving correct Chebychev approximations of real
number functions in the Coq theorem prover~\cite{coq}.
Also in Coq, Martin-Dorel and Melquiond~\cite{martin2016proving} verify
polynomial approximations using the CoqInterval~\cite{martin2016coqinterval}
package.
They develop a fully automated tactic for proving approximations inside
floating-point mathematical libraries correct.
A key difference between Dandelion and both these tools is that they cannot
certify approximations computed by Remez-like algorithms, which can in general
provide more accurate approximations.

For manual proofs, versions of the exponential function have been verified by
Harrison~\cite{HarrisonExponential}, and Akbarpour \etal{}~\cite{AkbarpourExponential}.
The manual proof by Harrison~\cite{harrison1997verifying} layed out the
foundations for Dandelion.
The work has also been extended by Chevillard \etal{}~\cite{chevillard2011}.
Instead of verifying approximation errors for polynomials, they use so-called
sum-of-squares decompositions~\cite{harrison-sos} to certify infinity norm
computations.
A major limiting factor for their work was finding accurate enough Taylor
polynomials which we found to not be a major issue for our approach.

Coward \etal{}~\cite{cowardformal} use the MetiTarski automated theorem
prover~\cite{akbarpour2010metitarski} to verify accuracy of hardware finite-precision
implementations of elementary functions.
MetiTarski provides proofs in machine-readable form using the TSTP
format~\cite{sutcliffe2004tstp} instead of being developed inside an
interactive theorem prover like HOL4.
%While as automated as Dandelion, MetiTarski does not provide rigorous
%machine-checked proofs.
A major conceptual difference is that the verification done by Coward \etal{}
reasons about bit-level accuracy of the hardware implementation, while Dandelion
reasons about real-number functions and polynomials.
Together with a verified roundoff error analysis like
FloVer~\cite{becker2018verified}, Dandelion could be extended to verify
finite-precision implementations of elementary functions, and together with
Daisy~\cite{izycheva2019synthesizing} verification could possibly be lifted to
entire arithmetic kernels.

A different style of unverified approximations is provided by Lim
\etal{}~\cite{lim2022one}.
Instead of computing specialized polynomial approximations, they focus
on correctly-rounded, general purpose approximations.
These approximations are not build for specific use-cases, but should
rather be seen as replacements for the functions provided in mathematical
libraries.
At the time of writing, their approach is not formally verified, but they
do provide a pen-and-paper correctness argument for their code generation.
The CR-libm~\cite{daramy2003cr} library also provides
unverified alternatives of correctly rounded mathematical
libraries, and Muller~\cite{muller2006elementary} gives a general
overview of the techniques for implementing elementary functions.
%and metalibm~\cite{kupriianova2014metalibm}
%\hb{Is this sufficient? I must admit I am not an expert on this area...}

\emph{(Automated) Real-Number Theorem Proving}
Dandelion heavily relies on HOL4's support for real-number theorem proving.
Below we list some alternatives for proving properties of real-numbers in both
interactive and automated theorem proving systems.
In the HOL-family of ITP systems, Harrison~\cite{harrison-sos} has formalized
sum-of-squares certificates for the HOL-Light~\cite{hollight} theorem prover.
His approach relies on semidefinite programming to find a decomposition of a
polynomial into a sum-of-squares polynomial.
%The technique he develops could be an alternative to the Sturm sequence based
%reasoning in Dandelion as it can also prove queries of the form
%$\forall x. x_{lo} \leq x \leq x_{hi} \rightarrow | p\;x | \leq \varepsilon$.
%The sum-of-squares decompositions relies on an external solver for finding the
%decomposition and the results need to be accurately reconstructed in HOL-Light.
%Dandelion instead computes all of its results in-logic, apart from the initial
%guess of the confidence intervals for zeros of a polynomial.
%\ed{I am not sure what to take from this relatively long discussion, especially
%because you said before that SOS had some issues that Sturm's sequence did not have.}
Both Isabelle/HOL and PVS have been independently extended with implementations
of Sturm sequences~\cite{EberlSturm, narkawicz2015formally, narkawicz2018decision}.
Their main focus is not on verification of polynomial approximations, they rather
use Sturm sequences to prove properties about roots of polynomials,
non-negativity, and monotonicity.

Previously we have already mentioned the MetiTarski automated theorem
prover~\cite{akbarpour2010metitarski}, as an example of an automated theorem prover
for real-numbered functions.
However, MetiTarski is not the only automated prover for real-numbered functions.
Real-numbered functions are also supported by \eg{} dReal~\cite{gao2013dreal},
and z3's SMT theory for real-numbers~\cite{moura2008z3}.
%%% Local Variables:
%%% mode: latex
%%% TeX-master: "../main"
%%% End:

%% file: sections/conclusion.tex
\section{Conclusion}

We have presented Dandelion, a verified and fully automated certificate
checker for polynomial approximations of elementary functions computed with
Remez-like algorithms.
Dandelion splits the validation task into two clearly separated phases:
The first phase replaces elementary functions by high-accuray Taylor series,
and the second phase uses Sturm's theorem and an external oracle to validate
the approximation error.
Our evaluation has shown that Dandelion certifies approximation errors
computed by an off-the-shelf Remez-like algorithm, and Dandelion also certifies
approximation errors for Chebyshev approximations.
%%% Local Variables:
%%% mode: latex
%%% TeX-master: "../main"
%%% End: